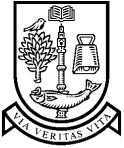

Department of Physics & Astronomy

Experimental Particle Physics Group
Kelvin Building, University of Glasgow,
Glasgow, G12 8QQ, Scotland.
Telephone: +44 - (0)141 – 3398855      Fax: +44 – (0)141 - 3305881

**UNIVERSITY
of
GLASGOW**

# Preliminary Results for LP VPE X-Ray Detectors.

R. Adams[+], R. Bates, C. Da Via, N.P. Johnson[*], V. O'Shea, A. Pickford, C. Raine and K. Smith.
Department of Physics & Astronomy, University of Glasgow, Glasgow G12 8QQ, Scotland.
* Dept. of Electronic & Electrical Engineering, University of Glasgow, Glasgow G12 8QQ, Scotland.
[+]Epitronics Corporation, 20012 North 19[th] Avenue, Phoenix AZ 85207, U.S.A.

*Abstract*

*Thick epitaxial layers have been grown using Low Pressure Vapour Phase Epitaxy techniques with low free carrier concentrations . This type of material is attractive as a medium for X-ray detection, because of its high conversion efficiency for X-rays in the medically interesting*

*energy range.*

# Preliminary Results for LP VPE X-Ray Detectors.

R. Adams[+], R. Bates, C. Da Via, N.P. Johnson*, V. O'Shea, A. Pickford, C. Raine and K. Smith.

Department of Physics & Astronomy, University of Glasgow, Glasgow G12 8QQ, Scotland.

* Dept. of Electronic & Electrical Engineering, University of Glasgow, Glasgow G12 8QQ, Scotland.

[+]Epitronics Corporation, 20012 North 19th Avenue, Phoenix AZ 85207, U.S.A.

## Abstract

*Thick epitaxial layers have been grown using Low Pressure Vapour Phase Epitaxy techniques with low free carrier concentrations . This type of material is attractive as a medium for X-ray detection because of its high conversion efficiency for X-rays in the medically interesting energy range.*

Very high growth rates have been achieved in vapour phase epitaxial growth systems by reducing the reactor pressure. The growth rate of some tens of microns per hour steadily decreases as the pressure is reduced and then begins to increase as the pressure is reduced still further. Growth rates of over 100 µm/hour have been achieved using this low pressure technique[1]. Thick epitaxial layers with a low enough free carrier concentration would be ideal as radiation detectors with particular advantages in the area of X-ray detection. Existing silicon based detectors have a low detection efficiency in comparison with GaAs as the photoelectric conversion probability scales with the fifth power of the atomic number of the detection medium. Even so, Si CCD's with a scintillating layer manage to compete with the most efficient film and screen systems in use to-day. Typically these still use only about 2% of the available radiation due to their low photon conversion efficiency.

As a first step in evaluating the potential of low pressure VPE growth for detectors, a test run to grow thick layers of high purity epitaxial GaAs was performed in a specially modified reactor at Epitronics Corp. 3 sample wafers were sent to Glasgow for material characterisation and detector evaluation. Local variations were enough to make the surface appear dull before polishing.

The layers were grown on 3 heavily doped 450 µm thick wafers supplied by Hitachi Cable. As detector characterisation with epitaxial layers grown on semi-insulating substrates is very difficult, the wafers were Si doped to $10^{18}$ cm$^{-3}$. The surface morphology of the grown layer showed rather large variations in wafer thickness with a strong gradient across it along one axis.

The thickness gradient is thought to be due to the gas flow across the wafer surface during growth and should be eliminated by appropriate modification of the glassware in the reactor. Because of the large gradient in layer thickness it was decided not to lap the epi-layer mechanically prior to processing which is part of our normal detector fabrication process. The wafer was chemically polished instead with a $H_2O_2:NH_4$ solution with pH maintained at 7.9 +/- 0.1 until roughly 30 µm of the layer was removed. It was hoped that this would give us optimum surface properties while removing a minimum of material from the layer. After polishing the small localised surface variations were no longer evident but the larger surface gradients were still present.

The wafer was processed to produce detector diodes of 3 mm diameter with 500 µm wide guard rings separated by a gap of 10 µm. The epi-layer contact was made with a Ti/Pd/Au Schottky contact.

The substrate contact used a Pd/Ge metallisation. A nitride encapsulation was deposited on both sides and apertures opened in the nitride layer to facilitate wire bonding.

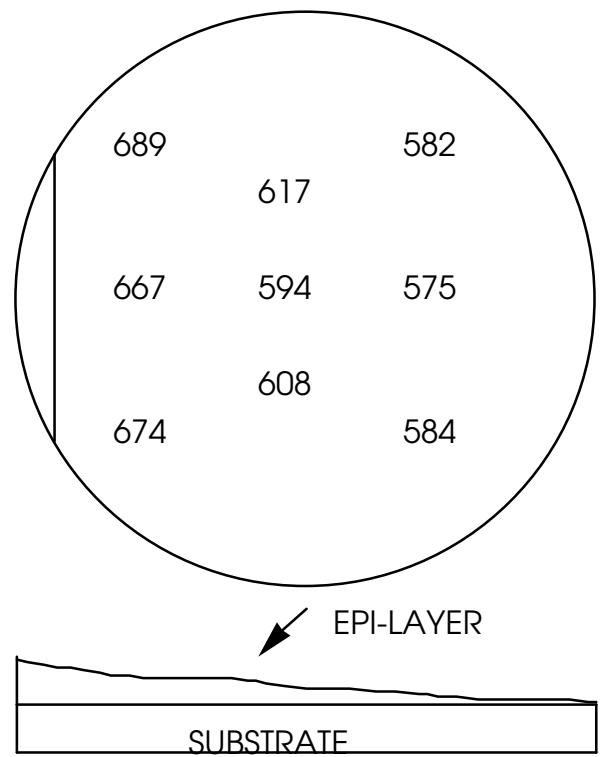

Fig[1] Measured wafer thickness in microns showing the gradient across the epi-layer. The unpolished surface had a dull appearance.

The diode characteristic shown in Fig[2] shows the leakage current of a typical detector from this batch. The thickness gradient across the wafer did not permit a good mask contact and all the guard rings on the detector structures were shorted to the detectors. The leakage current density is much more than 20 nA mm$^{-2}$ which is what we would expect for semi-insulating material[2] and much higher than expected for good diodes on epi material. This is due to a lack of adequate guarding of the structure which leads to leakage from the whole layer being measured instead of that due to the detector structure.

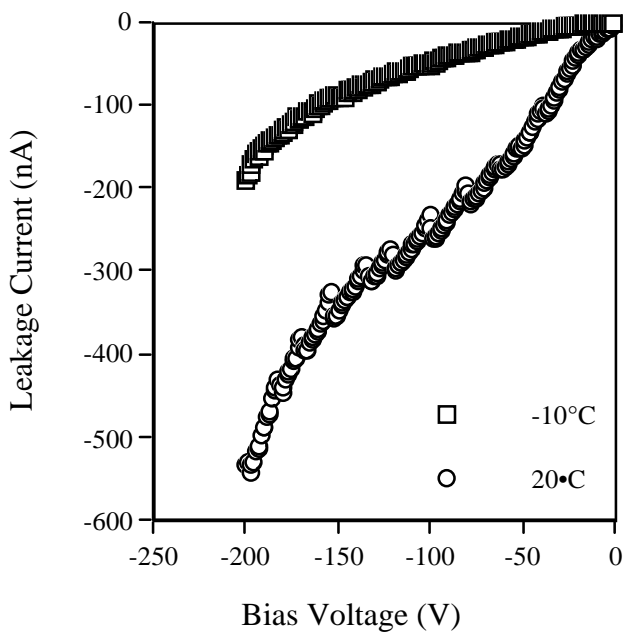

Fig[2]  3mm diameter detector leakage current characteristic at room temperature and -10 °C.

The capacitance of the detectors was measured at three frequencies to determine the free carrier concentration and also the depletion depth which corresponds to the active layer thickness of the detector. No variations in capacitance with frequency at low bias were observed. This is not the case with semi-insulating LEC material[3] and indicates that the trap concentration in this high growth rate epi-material is much lower than in LEC. DLTS measurements performed on this epi material indicate a concentration of $7 \times 10^{14}$ cm$^{-3}$ for the deep level EL2.

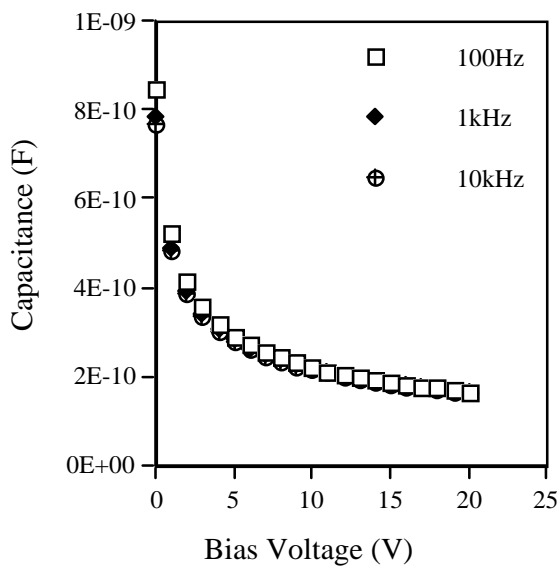

Fig[3] C-V measurement for 3mm diameter detector structures at low bias voltage.

The growth of the active thickness of the detector in LEC material is known to be a linear function of bias. The behaviour of depletion depth in the epi material shows the usual dependence on the square root of the bias. The plot of $1/C^2$ vs. bias shown in Fig[4] does not flatten off at any point indicating that the layer does not fully deplete at the highest bias that could be applied before breakdown.

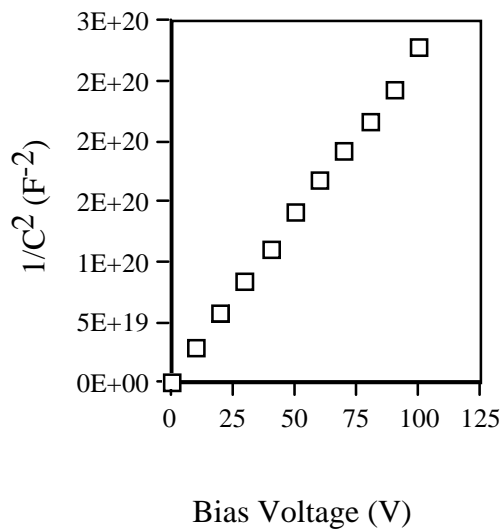

Fig[4] Full depletion is never reached due to the high free carrier concentration.

The measured capacitance was 73 pF at 100 V bias  and 42 pF for 200 V bias for the 3 mm pads. These give 20 μm and 27.5 μm respectively for the depletion layer thickness and show the epi-layer to have an effective free carrier concentration of around $3 \times 10^{14}$ cm$^{-3}$ for the diodes that we measured.

The diodes were also tested  as detectors to determine their charge collection efficiency. This is not 100% for LEC detectors but increases linearly with applied bias. The charge collection efficiency of the active layer was measured to be 100% for a variety of X-ray energies and the count rate for a given experimental set-up varied with the square root of the applied voltage. Fig[5] shows the measured spectrum from a 44 keV Terbium source. While the excess current in the present detectors limited the energies that could be resolved from noise to those above Ba (36 keV), it should not be a problem to resolve X-rays of much lower energies  with well-fabricated detectors.

## SUMMARY AND CONCLUSIONS

The grown layers are well suited to making high quality X-ray radiation detectors. The ability to grow thick layers of sufficiently high quality in reactors with capacities for many tens of wafers per run would make this the material of choice for most X-ray detection/imaging applications of the future. The biggest challenge to be overcome for this material is the reduction in free carrier concentration of the grown material as this determines the thickness of the active layer for any applied bias -- thick layers are only useful if they are fully depleted.

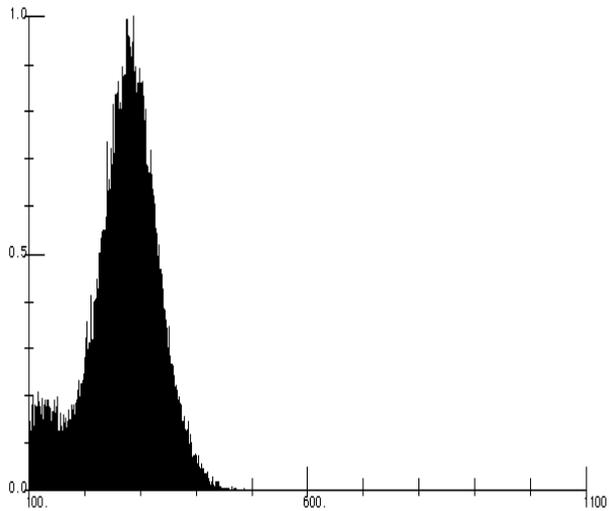

Fig[5] 44 keV Terbium X-ray spectrum with the epi-layer diode.

The difficulties in processing due to the surface properties of the wafers do not pose a serious problem. Mechanical polishing of the surface will remove most of the irregularities if the layer is grown sufficiently thick initially to produce a flat surface.

## REFERENCES

[1] K. Gruter et al., *Journal of Crystal Growth* **94** (1989) 607-612.

[2] S. Beaumont et al., *Nucl. Physics* B (Proc. Suppl.) **32** (1993) 296-299.

[3] S. Beaumont et al., *Nucl. Instr and Meth.* **A342** (1994) 83-89.